\documentclass[reprint,runinaddress,frontmatterverbose,twocolumn,showpacs,prl,floatfix,superscriptaddress]{revtex4-1}
\usepackage{graphicx}
\usepackage{dcolumn}
\usepackage{bm}
\usepackage{graphicx}
\usepackage{color}
\usepackage{colortbl}
\usepackage[dvipsnames]{xcolor} 
\usepackage{tabularx}
\usepackage{epsfig}
\usepackage{amsmath}
\usepackage{amssymb}
\usepackage{graphicx}
\usepackage{dcolumn,epstopdf}
\usepackage{bm}
\usepackage{wasysym}
\usepackage{times}
\definecolor{grn}{rgb}{0,0,0.54}

\usepackage[colorlinks,bookmarks=false,citecolor=blue,linkcolor=red,urlcolor=blue]{hyperref}
\usepackage{latexsym}
\usepackage{amssymb}
\usepackage{amsfonts}
\usepackage{amsmath}
\def\be{\begin{equation}}
\def\ee{\end{equation}}
\newcommand{\bea}{\begin{eqnarray}}
\newcommand{\eea}{\end{eqnarray}}

\begin{document}

\title{NMR study of the dynamics of $^3$He impurities in the proposed supersolid state of solid $^4$He}
\author{S. S. Kim}
\email{sungkim@phys.ufl.edu}
\author{C. Huan}
\author{L. Yin}
\author{J. Xia}
\affiliation{Department of Physics and National High Magnetic Field Laboratory, University of Florida, Florida 32611, USA}
\author{D. Candela}
\affiliation{Department of Physics, University of Massachusetts, Amherst, Massachusetts 01003, USA}
\author{N. S. Sullivan}
\affiliation{Department of Physics and National High Magnetic Field Laboratory, University of Florida, Florida 32611, USA}

\begin{abstract}
	The dynamics of $^3$He atoms in solid $^4$He have been investigated by measuring the NMR relaxation 
times $T_1, T_2$ in the region where a significant non-classical rotational inertia fraction (NCRIF) has been reported. 
	For $^3$He concentrations  $x_3$ = 16 ppm and 24 ppm, changes are observed for both the spin-lattice relaxation time $T_1$ and the spin-spin relaxation time $T_2$ at the temperatures corresponding to the onset of NCRIF and, at lower temperatures, to the $^3$He-$^4$He phase separation.
	The magnitudes of $T_1$ and $T_2$ at temperatures above the phase separation agree roughly with existing theory based on the tunneling of $^3$He impurities in the elastic strain field due to isotopic mismatch.
	However, a distinct peak in $T_1$ and a less well-resolved feature in $T_2$ are observed near the reported NCRIF onset temperature, in contrast to the temperature-independent relaxation times predicted by the tunneling theory.
\end{abstract}
\date{\today}
\maketitle

	The discovery of a non-classical rotational inertia fraction (NCRIF) in solid $^4$He by Kim and Chan
 \cite{ISI:000188068100037, ISI:000224136000041} 
has generated enormous interest because the NCRIF could be the signature of a supersolid state \cite{PhysRevLett.25.1543}.
	Several independent experiments \cite{PhysRevLett.100.065301, PhysRevLett.97.165301} have shown that the NCRIF magnitude and temperature dependence are strongly dependent on defects such as $^3$He impurities and the quality of the crystals.
	Recently studies of the elastic properties of solid $^4$He by Beamish and colleagues \cite{PhysRevLett.104.195301} have revealed a significant frequency dependent change in the elastic shear modulus with an enhanced dissipation peak having a temperature dependence comparable to that observed for the NCRIF. 
	These results suggest that the dynamics of the $^4$He lattice plays an important role in the low temperature bulk properties of solid $^4$He and rather than observing a phase transition to a supersolid state one may be observing a thermally excited dynamical response. 
	It is therefore important to study the microscopic dynamics of $^3$He impurities to better understand their role in the NCRIF and shear-modulus phenomena.
	To meet this need we have measured the NMR relaxation times of dilute $^3$He impurities in solid $^4$He at low temperature.
	The NMR relaxation rates are determined by quantum tunneling ({\it via} $^3$He-$^4$He atom exchange) and scattering of the diffusing atoms by the crystal deformation field around the $^3$He impurities and other lattice defects. 
	The NMR relaxation rates are therefore very sensitive to the elastic properties of the solid $^4$He and on any changes in the crystal ground state that would modify the tunneling rate.
	Although low-temperature NMR data for $^3$He impurities in solid $^4$He have been reported in Refs. \onlinecite{PhysRevB.81.214515, 2010JLTP...158..584} for higher $^3$He concentrations $x_3 \ge 100$~ppm, we report the first NMR data on isolated $^3$He impurities in the region of $x_3$ and $T$ in which NCRIF has clearly been observed.
	(Recently Toda {\it et al.} \cite{springerlink:10.1007/s10909-010-0279-z} reported simultaneous NCRIF and NMR data for a sample with $x_3 = 10$~ppm, but the NMR signal was only observable in their study from $^3$He atoms in phase-separated clusters.)
 
\begin{figure}[!hb]
\includegraphics*[width=0.32\textwidth]{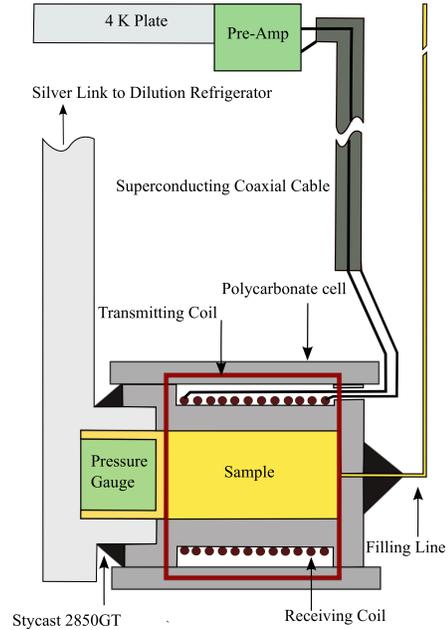}
\caption{ \label{fig: NMR cell design}
	(Color online) Schematic representation of the low temperature NMR cell.
	The preamplifier and tuning capacitor are located on a 4~K cold plate located a distance of 1.8~m from the sample cell.
	The RF transmitting and receiving coils are simplified in this figure.}
\end{figure}

\begin{figure}[!ht]
\includegraphics[width=1.1\linewidth]{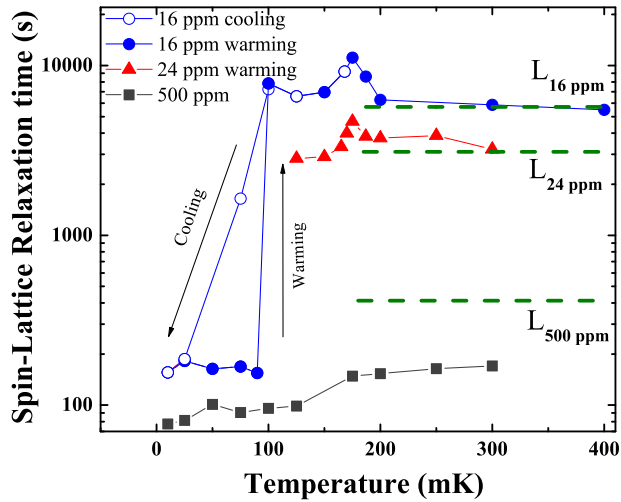}
\caption{\label{fig: T1}
	(Color online) Temperature dependence of the nuclear spin-lattice relaxation time $T_1$ for samples with $x_3$ = 16 ppm and 24 ppm compared to a high concentration sample ($x_3$ = 500 ppm, from Ref. \onlinecite{2010JLTP...158..584}).
	A peak in $T_1$ is observed for both samples at $T \approx 175$~mK
	In addition, for the $x_3$ = 16 ppm sample $T_1$ drops by a factor of $\sim 100$ below 85 mK due to $^3$He-$^4$He phase separation.
	Solid (open) circles show data taken on warming (cooling). 
	The dashed green lines represent $T_1$ values calculated for the impuriton model \cite{Landesman1975137}.}
\end{figure}

\begin{figure}[ht]
\includegraphics[width=1.1\linewidth]{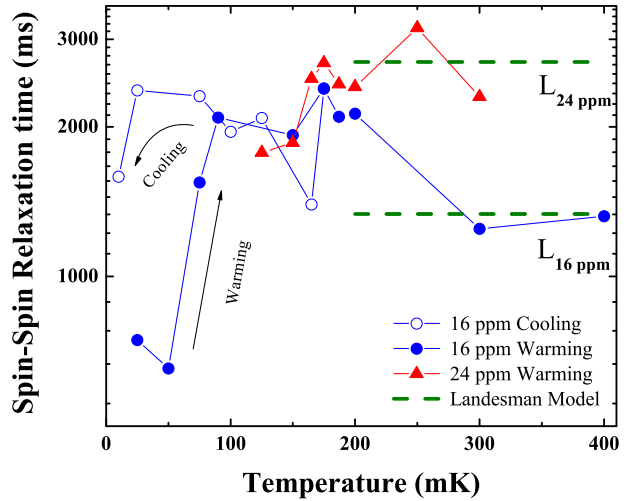}
\caption{\label{fig: T2}
	(Color online) Temperature dependence of the nuclear spin-spin relaxation time $T_2$ for samples with $x_3$ = 16 ppm and 24 ppm.
	In addition to a strong, hysteretic drop in $T_2$  for $x_3 = 16$~ppm below 100~mK due to phase separation, both samples show poorly-resolved features near the temperature at which the $T_1$ peaks are observed, $T \approx 175$~mK.
	As $T_2$ reflects the spectral density of $^3$He motion near zero frequency it is sensitive to slow and static $^3$He redistributions that do not affect $T_1$.
	This may also contribute to the temperature dependence of the $T_2$ values observed at $T > 200$~mK for the $x_3 = 24$~ppm sample, which was cooled after less annealing than the $x_3 = 16$~ppm sample.}
\end{figure}

	The samples were prepared by mixing high purity gases and condensing the mixture at high pressure (46.2 bar) into a polycarbonate cell that contained a pressure gauge (Fig.~\ref{fig: NMR cell design}), and then solidifying the samples using the blocked capillary method.
	Thermal contact to the sample was provided by a solid silver cold finger extending from a dilution refrigerator.
	Details of the cell have been reported elsewhere \cite{2010JLTP...158..584}.
	A first sample with $x_3 = 16$~ppm was annealed for 24 hours just below the melting point, while a second sample with $x_3 = 24$~ppm was annealed for only 30 minutes.
	For both samples the final pressure measured {\it in situ} at low temperature was $27.75(5)$~bar, corresponding to a molar volume $v_\text{m} \approx 20.8$~cm$^3$. 

	Standard pulsed NMR techniques were used to measure the nuclear spin relaxation times: magnetization recovery following a spin echo to measure $T_1$, and a Carr-Purcell-Meiboom-Gill \cite{PhysRev.94.630} multiple-echo sequence to measure $T_2$.
	Separate orthogonal coils were used for excitation and signal reception to minimize sample heating and pulse pickup by the cryogenic preamplifier.
	The coils were isolated from the sample cell and allowed to float independently from the sample temperature which was monitored using a carbon resistor calibrated against a $^3$He melting curve thermometer.
	The NMR signals at these concentrations are very weak and to improve the signal-to-noise ratio we used a low temperature RF preamplifier that could operate in a strong applied magnetic field \cite{2010JLTP...158..692}. 
	Even with this preamplifier signal averaging up to five days per data point was needed to realize an adequate signal-to-noise ratio.

	The studies were carried out for a Larmor frequency of $\omega_\text{L}/2\pi = 2$ MHz as the expected relaxation times extrapolated from previous studies \cite{ PhysRevLett.29.708, 1984JLTP...13..65, PhysRevLett.34.1545, 1984JLTP...57..179} would be prohibitively long at higher frequencies.
	In order to determine the $^3$He concentrations of the samples we measured the amplitude of the NMR echoes at high temperatures ($150<T<350$ mK) 
which follows the Curie law and compared the signal to a standard reference sample with $x_3 = 1000$~ppm.

	The observed temperature dependences of the nuclear spin-lattice relaxation time $T_1$ for samples with $x_3 = 16$~ppm and 24~ppm are compared with the dependence for a much higher concentration sample ($x_3 = 500$~ppm from Ref.~\onlinecite{2010JLTP...158..584}) in Fig.~\ref{fig: T1}.
	A pronounced peak at $T \approx 175$~mK is observed for both the 16 ppm and 24 ppm samples in contrast with the weak temperature dependence observed for the $x_3 = 500$~ppm sample.
	Samples with higher concentrations ($500 \text{~ppm} < x_3 \leq 2000$~ppm) were also studied \cite{2010JLTP...158..584, springerlink:10.1007/s10909-010-0260-x} and had temperature dependences similar to that of the 500~ppm sample except for the expected shift in the phase separation temperature.
	The position of the peak in $T_1$ corresponds closely to the saturation temperature $T_{90}$ observed for NCRIF at this $x_3$ \cite{PhysRevLett.100.065301}.
	Allowing for expected shifts with pressure and $x_3$, the $T_1$ peak we observe also correlates well with the shear dissipation peak observed by Syschenko {\it et al.} \cite{PhysRevLett.104.195301}.
	Finally, the $T_1$ peak occurs at approximately the same temperature for this $x_3$ value as the sharp ultrasonic absorption anomaly reported by by Ho {\it et al}. \cite{springerlink:10.1007/s10909-005-0094-0}.

	In parallel with the observations for $T_1$, a less well resolved peak in $T_2$ is observed for $x_3 = 16$~ppm at the same temperature for which the $T_1$ peak is seen (Fig.~\ref{fig: T2}).
	At lower temperatures large changes in $T_1$ and $T_2$ are observed due to $^3$He-$^4$He phase separation at $T_\text{ps} \approx 85$~mK.
	This value of $T_\text{ps}$ agrees well with the predictions of Edwards and Balibar \cite{PhysRevB.39.4083}. 
	The same features are observed for the sample with 24~ppm: a strong peak in $T_1$ at $T \approx 175$~mK (Fig.~\ref{fig: T1}) and a weaker peak in $T_2$ (Fig.~\ref{fig: T2}).
	The $T_1$ and $T_2$ features at $T \approx 175$~mK appear unrelated to the large changes in $T_1$ and $T_2$ associated with the phase separation at lower temperatures. 
	To verify this fact, the 24 ppm sample was purposely never cooled to the phase-separation temperature.

	The thermal hysteresis around 85 mK in $T_1$ and $T_2$ reinforces the view that the changes at this temperature are due to $^3$He-$^4$He phase separation \cite{1972JLTP...8..3}.
	The NMR echo amplitudes become temperature independent below the phase separation as expected for the formation of degenerate liquid $^3$He droplets \cite{springerlink:10.1007/s10909-010-0260-x}.
	Our results for the phase separation temperatures are in good agreement with those reported by Toda {\it et al}. for $x_3>100$ ppm \cite{PhysRevB.81.214515}.

	The samples with $x_3 = 16$~ppm and $x_3 = 24$~ppm have significantly different temperature dependences for $T_2$ at $T > 200$~mK.
	As $T_2$ (unlike $T_1$) is sensitive to very slow and static changes in the positions of the $^3$He impurities, it seems that the $T_2$ data cannot clearly resolve the effects leading to the $T_1$ peak from other effects.
	In particular, the higher-temperature data in Fig.~\ref{fig: T2} suggest that $T_2$ is sensitive to crystal quality.

	The nuclear spin dynamics of $^3$He impurities in solid $^4$He have been studied extensively for relatively high concentrations ($x_3>90$ ppm) and high temperatures ($T>350$ mK) \cite{PhysRevLett.29.708, 1984JLTP...13..65, PhysRevLett.34.1545, 1984JLTP...57..179}.
	The results have been described in terms of mobile $^3$He impurities tunneling through the $^4$He matrix by $^3$He-$^4$He exchange ($J_{34}$) with a mutual scattering due to the elastic deformation field surrounding each impurity ($K(r)=K_0r^{-3}$) \cite{RevModPhys.43.532, Landesman1975137}. 
	A reasonable fit to the $T_1$ and $T_2$ values observed for very low concentrations ($\sim 20$ ppm) at high temperature ($T > 200$~mK) is obtained using the Landesman model \cite{Landesman1975137} as shown by dashed green lines in Figs. \ref{fig: T1} and \ref{fig: T2}. 
	The $^3$He atoms can also tunnel as weakly bound pairs \cite{PhysRevLett.35.1007}.
	The pair-tunneling model explains resonant dips in $T_1$ observed \cite{PhysRevLett.34.1545} at $\omega_\text{L} = 1.3$ and 2.6~MHz since $T_1$ directly measures fluctuations in $^3$He-$^4$He separations at frequencies of $\omega_\text{L}$ and $2\omega_\text{L}$.
	However, none of these models predict temperature dependent relaxation times so they can not explain the observed $T_1$ peaks.

\begin{figure}[!Hb]
\includegraphics*[width=0.53\textwidth]{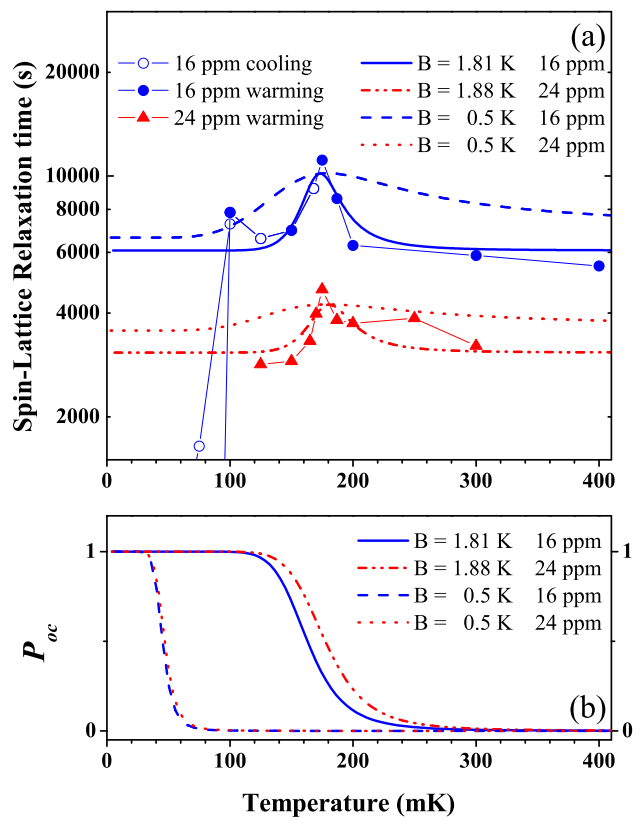}
\caption{\label{fig: T1fits}
    (Color online) (a) Measured spin-lattice relaxation times for two low-$x_3$ samples (data points) along with fits to a phenomenological model based on thermally-activated relaxation of unknown degrees of freedom (smooth curves). 
	(b) Temperature-dependent probability for a defect such as a dislocation to be occupied by a $^3$He impurity.
	Here $B$ is the $^3$He thermal activation energy (a) or defect binding energy (b).}
\end{figure}

	The $T_1$ peak occurs at roughly the same temperature at which torsional oscillator and shear modulus anomalies are observed \cite{PhysRevLett.100.065301, PhysRevLett.104.195301}, and it is tempting to infer a connection between all three phenomena.
 	One possibility is that the sharp reentrant peak in fluctuations near 175 mK signals a phase transition, possibly associated with supersolidity.
	However, other explanations not involving a phase transition must be considered.
	As shown in Ref. \onlinecite{PhysRevLett.100.065301}, the $x_3$-dependent 
onset temperature for the torsional-oscillator anomalies agree well with the temperature $T_\text{IP}(x_3)$ below which $^3$He pinning of dislocation lines is expected to dominate over pinning by dislocation-network nodes, $T_\text{IP} = B/\ln(a/x_3L_\text{N})$ where $B\approx 0.5$ K is the binding energy of a $^3$He impurity to a dislocation line, $L_\text{N}$ is the average length of dislocation segments between nodes, and $a = 3.7\times10^{-10}$~m is the nearest-neighbor distance.
	In this model there are roughly $L_\text{N}/a$ binding sites for a $^3$He impurity on an internode segment of a dislocation line, and each binding site is occupied with probability $P_\text{oc} = (e^{-B/T}/x_3+1)^{-1}$.

	For a total length of dislocation lines per unit volume $\Lambda \approx 0.2/L_\text{N}^2 \approx 10^{11}$~m$^{-2}$ for $^4$He crystals grown by the blocked-capillary method \cite{PhysRevLett.100.065301}, the concentration of $^3$He binding sites on dislocations relative to the number of $^4$He atoms in the sample is $x_\text{d} \approx 10^{-8}$, much smaller than the concentrations of $^3$He atoms $x_3$ present in our NMR experiments.
	It would be surprising if such a small concentration of binding sites $x_\text{d}$ could have a measurable effect on $T_1$.
	However, there are other indications that the density of dislocations or other defects must be much larger than the estimate quoted above, if the NCRIF observations are explained either by superflow along dislocations cores \cite{PhysRevLett.99.135302} or as a direct mechanical effect unconnected with supersolidity \cite{PhysRevLett.104.255301}.

	Figure \ref{fig: T1fits} shows the temperature dependence of $T_1$ along with the occupation probability for $^3$He binding sites and another model discussed below.
	With the expected $^3$He-dislocation binding energy $B \approx 0.5$ K it can be seen that the significant transition in $P_\text{oc}(T)$ occurs at much lower temperatures than the peak in $T_1$ as shown in Fig. \ref{fig: T1fits}(b). 
	If larger binding energies $B = 1.8 - 1.9$~K are assumed, the drop in $P_\text{oc}(T)$ occurs at temperatures close to those of the NMR anomaly.
	However, it is not clear how the partial occupation of $^3$He binding sites would lead to reduced fluctuations of $^3$He interatomic vectors as inferred from the peak in $T_1$.
	
Another phenomenological model we consider is to associate the $T_1$ anomalies with a thermally activated relaxation peak as has been used successfully to describe the shear-modulus shifts \cite{PhysRevLett.104.195301}.
	The smooth curves in Fig. \ref{fig: T1fits}(a) are fits to the form $T_1 = [R_0 - 2R_1\omega\tau/(1+(\omega\tau)^2)]^{-1}$, $\omega\tau = \omega\tau_0 e^{B/T}$.
	Here $R_0$ and $R_1$ are fitting parameters giving the background relaxation rate and height of the peak in $T_1$, $B$ is the activation energy, $\tau_0$ is an attempt frequency, and $\omega$ is the frequency at which relaxation is being probed 
(e.g. $\omega_\text{L} = 1.26 \times 10^7 s^{-1}$).
	For the dotted and dashed curves the activation energy was fixed at $B = 0.5$~K (the approximate binding energy of a $^3$He impurity to a dislocation inferred in Ref. \cite{PhysRevLett.100.065301}) while for the other curves 
$B$ was allowed to increase to improve the fit to the data.
	Note that these fits determine only the combination $\omega\tau_0$ (not $\omega$ or $\tau_0$ separately); for the best fit to the $x_3 = 16$~ppm 
data $\omega\tau_0 = 2.9\times10^{-5}$ and $B = 1.81$ K.

	It must be emphasized that the fitting functions used in Fig. \ref{fig: T1fits}(a) are phenomenological and not derived from a simple microscopic model of the NMR relaxation.
	For example, a simple resonance between the thermally-activated motion of $^3$He impurities and the Larmor frequency $\omega_\text{L}$ would ordinarily lead to a {\it minimum} in $T_1$ (increased fluctuations at the Larmor frequency), not a peak as we observe.
	On the other hand, relaxation of $^3$He impurities in solid $^4$He is known to be controlled by a nonmonotonic spectral density of fluctuations with a sharp feature at the frequency ($\sim$ 3 MHz) at which nearest-neighbor $^3$He pairs ``walk'' through the lattice via quantum tunneling \cite{PhysRevLett.35.1007}.
	Therefore a more complex mechanism such as the disruption of the 
quantum walking motion of $^3$He pairs by resonant fluctuations of dislocation lines might be needed to explain our NMR data.

        The NMR measurements can only be understood in terms of a sharp change in the fluctuation spectrum and the same changes would also play a dominant role in the anomalies observed for the sound attenuation and the shear modulus.
	Specifically, if the fluctuation spectrum is associated with critical behavior at a phase transition, the attenuation and NMR relaxation rates would vary as $T_1\sim\ |T-T_0 |^\lambda$ in agreement with the NMR and sound attenuation results.
	Alternatively, if a collective but non-critical change occurred in the lattice dynamics, the associated change in the elastic properties of the solid would result in changes of both the observed shear modulus and the NMR relaxation rates. The latter has been shown to depend on the elastic strain surrounding $^3$He impurities \cite{Landesman1975137}.
	Further studies for a wide frequency range and for samples grown at constant pressure to produce higher quality crystals are needed to distinguish between these interpretations.
	However, it is clear that a successful microscopic model of the lattice dynamics of solid $^4$He must explain not only the NCRIF observations, but also the coincident anomalies that have been observed in ultrasound \cite{springerlink:10.1007/s10909-005-0094-0}, shear modulus \cite{PhysRevLett.104.195301}, and NMR relaxation as reported here.

\begin{acknowledgments}
	This research was carried out at the NHMFL High B/T Facility which is supported by NSF Grant DMR 0654118 and by the State of Florida.
	This project was supported in part by an award from the Collaborative Users Grant Program of the NHMFL.
	We gratefully acknowledge useful discussions with Alan Dorsey, Pradeep Kumar, Moses Chan and Brian Cowan.
\end{acknowledgments}

\bibliography{2011-02-09_3He}

\end{document}